\documentclass[10pt]{article}
\pdfoutput=1
\usepackage[ruled,vlined]{algorithm2e}
\usepackage{multirow}
\usepackage{hyperref}
\usepackage{graphicx}

\title{Structural representations of DNA regulatory substrates can enhance sequence-based algorithms by associating functional sequence variants}

\author{
  Jan Zrimec \\
  Chalmers University of Technology\\
  Gothenburg, Sweden \\
  \texttt{janzrimec@gmail.com} \\
}

\date{13 June 2020}

\begin{document}
\maketitle

\begin{abstract}
  The nucleotide sequence representation of DNA can be inadequate for resolving protein-DNA binding sites and regulatory substrates, such as those involved in gene expression and horizontal gene transfer. Considering that sequence-like representations are algorithmically very useful, here we fused over 60 currently available DNA physicochemical and conformational variables into compact structural representations that can encode single DNA binding sites to whole regulatory regions. We find that the main structural components reflect key properties of protein-DNA interactions and can be condensed to the amount of information found in a single nucleotide position. The most accurate structural representations compress functional DNA sequence variants by 30\% to 50\%, as each instance encodes from tens to thousands of sequences. We show that a structural distance function discriminates among groups of DNA substrates more accurately than nucleotide sequence-based metrics. As this opens up a variety of implementation possibilities, we develop and test a distance-based alignment algorithm, demonstrating the potential of using the structural representations to enhance sequence-based algorithms. Due to the bias of most current bioinformatic methods to nucleotide sequence representations, it is possible that considerable performance increases might still be achievable with such solutions.
\end{abstract}


\section{Introduction}
Besides the popular yet simplistic representation of DNA as a polymer chain of 4 different nucleotide bases A, C, G and T, the molecule in its double stranded dsDNA form possesses certain conformational and physicochemical properties. These are especially important in relation to interactions of the DNA with proteins, which drive many essential cellular processes \cite{Rohs2009-hm,Zrimec2015-xf,Zrimec2018-lx}. These include but are not limited to transcription \cite{Rohs2009-hm}, replication \cite{Chen2012-gd} as well as horizontal gene transfer (HGT) \cite{Zrimec2018-lx,Zrimec2020-wx}. The processes are commonly initiated and regulated at specific DNA regulatory regions, which are the substrates for protein binding and enzymatic activity, and include promoters \cite{Watson2008-dt}, origins of replication \cite{Chen2012-gd} and origins of transfer (transfer regions, plasmid conjugation in HGT) \cite{De_La_Cruz2010-xj}. The DNA substrates contain either one or multiple protein binding sites, such as for instance transcription factor binding sites (TFBS) in promoters, as well as additional sequence and structure context that is related to protein-DNA recognition and binding preceding the main enzymatic processing \cite{Levo2015-iu,Marcovitz2013-kg,Zrimec2018-lx}.

As such, the nucleotide sequence representation is often not sufficiently detailed for discriminating DNA regulatory substrates, as it encodes specific conserved structural properties that are not directly obvious from the mere sequence context (Table 1) \cite{Zrimec2018-lx,Tosato2017-nq,Slattery2014-ne}. To uncover the encoded structural properties, many DNA structure models and prediction tools have been developed, including (i) DNA thermodynamic stability and its potential for destabilization and melting bubble formation, modelled with the nearest neighbor (NN) framework \cite{SantaLucia1998-hc} and thermally induced duplex destabilization \cite{Zrimec2015-xf}, (ii) DNA major and minor groove properties that describe their size and thus accessibility by proteins, which are the focus of the DNAshape and ORChID2 models \cite{Bishop2011-jm,Chiu2016-kb} among others, (iii) DNA intrinsic curvature and flexibility, modeled by measuring DNAzeI digestion profiles \cite{Brukner1995-pt} and DNA persistence lengths \cite{Geggier2010-mw}, (vi) DNA twisting and supercoiling, captured by multiple DNA conformational models and variables \cite{Karas1996-qz,Olson1998-rw,Perez2004-sx}, (v) differences in DNA spacing and orientation between binding and enzymatic sites, described for instance by DNA helical repeats \cite{Geggier2010-mw}, and (vi) the propensity for transitions between DNA forms, such as from B-DNA to A-DNA or to Z-DNA given with respective variables \cite{Aida1988-iq,Hartmann1989-ji,Kulkarni2013-xm,Ho1986-hg}. However, due to the large variety and differences among the DNA structural models and variables, it is not simple to choose among them or integrate them within existing DNA bioinformatic frameworks. Current studies thus focus on specific groups of DNA structural properties and do not span the whole possible structural repertoire \cite{Chiu2016-kb,Zrimec2018-lx,Zrimec2013-ds,Samee2019-xj,Bansal2014-ko,Chen2012-gd}. A common DNA structural representation could thus help circumvent these problems and facilitate the development of novel, improved DNA algorithms. 

\begin{table*}
\footnotesize
  \caption{Overview of DNA structural properties and representative variables in protein-DNA interaction.}
  \begin{tabular}{p{3cm}|p{3cm}|p{3cm}|p{1cm}}
    \hline
    DNA structural properties & Facilitated protein-DNA interactions & Representative structural variables & References\\
    DNA stability, destabilizations and melting bubble formation & Enzymatic processing of substrates, e.g. nicking of transfer regions, leads to secondary structure formation & Duplex stability, Thermally induced duplex destabilization (TIDD) & \cite{SantaLucia1998-hc,Lucas2010-gi,Zrimec2015-xf,Sut2009-kg}\\
    \hline
    Major and minor groove properties & Readout of chemical information, e.g. TFs in promoters & DNAShape, ORChID2 & \cite{Rohs2009-hm,Chiu2016-kb,Bishop2011-jm,Watson2008-dt}\\
    \hline
    Intrinsically curved or flexible regions & Binding and topological changes, e.g. IHF binding in promoters & DNAzeI cleavage frequency, Persistence length & \cite{Brukner1995-pt,Geggier2010-mw,Moncalian1999-qj}\\
    \hline
    DNA twist and supercoiling & Topological changes recognized by proteins, e.g. histones, and affect other properties & Twist and other conformational variables & \cite{Karas1996-qz,Olson1998-rw,Perez2004-sx,Watson2008-dt}\\
    \hline
    Differences in DNA spacing and orientation in binding and enzymatic sites & Affect binding with multiple contact points and protein complex formation & Helical repeats & \cite{Williams2007-be,Geggier2010-mw,Watson2008-dt}\\
    \hline
    Propensity for transitions between DNA forms B-DNA, A-DNA, Z-DNA & Affect overall features recognized by proteins and their accessibility & B-A and B-Z transition propensities & \cite{Aida1988-iq,Ho1986-hg,Hartmann1989-ji,Kulkarni2013-xm}\\
    \hline
  \end{tabular}
\end{table*}

Hence, the aim of the present study was to analyse and engineer novel DNA structural representations for use with bioinformatic frameworks, such as sequence alignment and motif finding algorithms, and to compare them with the standard nucleotide sequence-based methods. First, based on DNA physicochemical and conformational properties that are involved in DNA-protein interactions and using dimensionality reduction techniques we constructed different structural DNA representations. To explore the smallest amount of structural information that could sufficiently describe functional DNA regions we compressed the DNA representations using clustering algorithms to encompass from an estimated 2 to 8 bits of information. Next, we explored the capability of each representation to encode multiple DNA sequence variants using TFBS motif datasets. To facilitate the usefulness of the representations with existing DNA algorithms via comparison of the (dis)similarity of encoded sequences, we developed a structural distance function and tested it on datasets of transfer and promoter regions. Finally, we explored the applicability of the structural representations within a sequence alignment framework and discussed ideas for hybrid sequence and structure-based approaches specifically for analysing regulatory DNA substrates.

\section{Methods}
\subsection{Datasets}
We obtained 64 published DNA structure models that were shown to be informative for analysis of DNA-protein interactions \cite{Zrimec2020-wx} (Table 1). These models were based on nearest neighbor dinucleotide (56), trinucleotide (4) and pentanucleotide (4) models and included physicochemical and conformational properties and properties attributed to DNA-protein interactions. Of these, 44 models were derived experimentally and the rest computationally based on experimental data. The set included the widely used models DNA shape \cite{Rohs2009-hm,Chiu2016-kb}, Orchid \cite{Bishop2011-jm}, DNA stability and thermally induced duplex destabilization \cite{SantaLucia1998-hc,Zrimec2015-xf}. 

A dataset of transcription factor binding site (TFBS) motifs was obtained from the Jaspar database \cite{Khan2018-wj} (sites file) and filtered to contain only sequences with \{A,C,G,T\} characters, of equal length as the median length in each motif group but at least 9 bp long and containing at least 10 motif sequence variants.

A published dataset of transfer regions from 4 mobility (Mob) groups \cite{Zrimec2018-lx} was used as positive examples and expanded with 64 negative examples. Negative example sequences were selected randomly from a region 200 to 800 bp around the enzymatic nicking sites \cite{Zrimec2018-lx}, thus containing different non-regulatory coding and non-coding regions, and low sequence similarity was verified among the sequences (p-distance \textgreater{} 0.6). The part of the transfer regions with relevant protein binding features from -140 bp to +80 bp according to the nicking site was used \cite{Zrimec2018-lx}. For testing the alignment algorithm, a second published dataset of 112 transfer regions (queries) of equal length as the ones above and 52 plasmids (targets) \cite{Zrimec2020-wx} from 4 Mob groups was used.

A dataset of Escherichia coli promoter regions was obtained \cite{Gusmao2014-hp} with 100 bp positive and negative examples (positive, mixed1 and control). We randomly sampled 200 sequences from each dataset to create a 600 element dataset. The second dataset of Escherichia coli promoter regions was obtained from Regulon DB v9 \cite{Gama-Castro2016-so} and contained only 81 bp positive examples grouped according to 6 sigma factors \cite{Watson2008-dt}. A random sampling of 94 sequences (size of smallest group) from each group yielded a dataset with 564 elements.

\subsection{Construction and analysis of DNA structural representations}
To develop DNA structural representations we computed the structural properties of all permutations of k-mers 3, 5, 7 and 9 bp in length (1 to 4 neighboring regions around a specific nucleotide) and performed dimensionality reduction followed by clustering (Fig. 1A). The structurally defined groups of k-mers were termed s-mers. Structural properties were calculated in windows of 5 bp or at default values as described \cite{Zrimec2015-xf,Zrimec2018-lx}. Dimensionality reduction and analysis of the main components of variance was performed using Principal Component Analysis (PCA). The k-means clustering algorithm was used (Matlab), where clusters with a lowest total sum of distances were chosen from 10 runs of up to 1000 iterations at default settings. The optimal amount of clusters was analysed with the Elbow, average Silhouette \cite{Rousseeuw1987-mx} and GAP \cite{Tibshirani2001-wq} methods with Matlab function evalclusters at default settings with triplicate runs. The tested numbers of clusters included 4, 8, 16, 32, 64, 128 and 256 clusters, chosen considering that (i) positions with $2^x$ possible states (clusters) can carry a maximum of x bits of information \cite{Schneider1986-iv}, (ii) 1 bp of DNA carries up to 2 bits of information (iii) up to 4 bp neighboring regions defined the structural effect in the s-mers, and (iv) s-mers are overlapping, meaning information is distributed among all of them. Thus, up to 8 bits of information (256 clusters) was expected per s-mer (and less with decreasing s-mer size). 

\begin{figure}[ht]
  \centering
  \includegraphics[width=7.5cm,keepaspectratio]{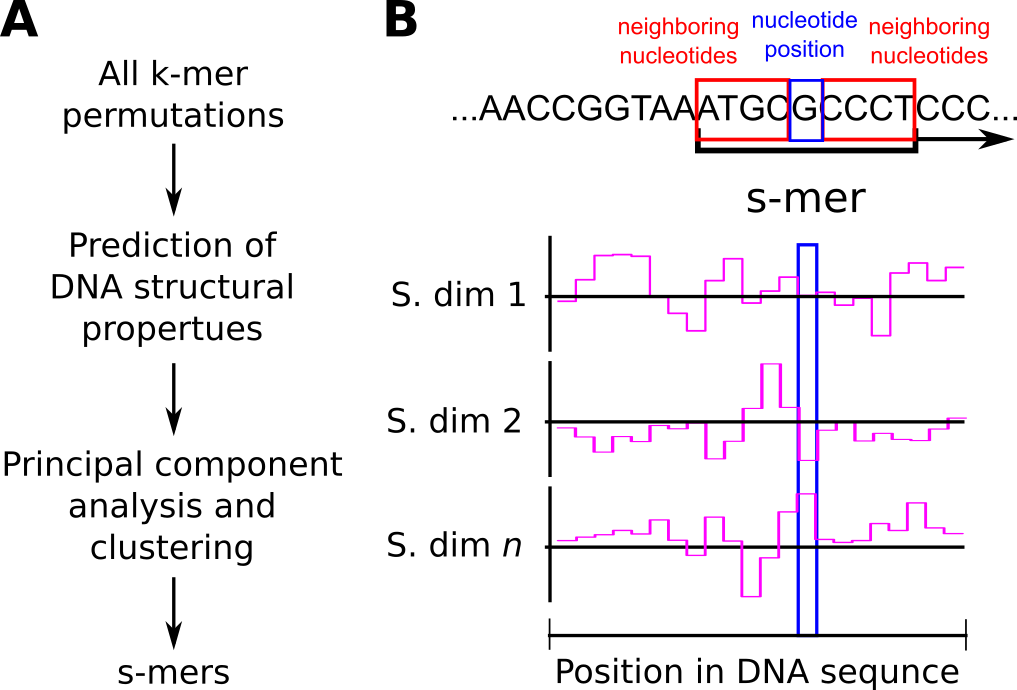}
  \caption{Schematic depiction of the (A) construction and (B) usage of structural representations. In a structural representation of a given DNA sequence, each central nucleotide position and its neighboring regions define a k-mer from 3 to 9 bp in length, and are encoded as an s-mer with n structural dimensions (S. dim.) that can be defined as a sequence of s-mer cluster centroids.}
\end{figure}

For a DNA substrate, the length of the structural representation was equal to the length of the nucleotide sequence minus the leftover nucleotides at the borders equal to $(s-1)/2$, due to the neighboring nucleotides in s-mers (Fig. 1B). The s-distance between two DNA substrates was the sum of squared Euclidean distances between the cluster centroids of all equally positioned s-mers in their structural representations of length \textit{n},
\begin{equation}
  s-distance=\sum_{i=1}^{n}d(C_{1i}, C_{2i})^2,
\end{equation}
where \(C_{1i} = (c_{11}, c_{12},..., c_{1k})\) is the cluster centroid of the s-mer at position \textit{i} of the first sequence and \(C_{2i} = (c_{21}, c_{22},..., c_{2k})\) is the cluster centroid of s-mer at position \textit{i} of the second sequence.

\subsection{Statistical analysis and performance metrics}
For statistical hypothesis testing, the Python package Scipy v1.1.0 was used with default settings. To evaluate the explained variation, the coefficient of determination was defined as
\begin{equation}
R^2=1-\frac{SS_{Residual}}{SS_{Total}},
\end{equation}
where \(SS_{Residual}\) is the within group residual sum of squares and \(SS_{Total}\) is the total sum of squares. The compression ratio was calculated as
\begin{equation}
Compression\: ratio = \frac{\textrm{Num. unique k-mers}}{\textrm{Num. unique s-mers}}.
\end{equation}
Precision and recall were defined as 
\begin{equation}
Precision = \frac{TP}{TP+FP},
\end{equation}
\begin{equation}
Recall = \frac{TP}{TP+FN},
\end{equation}
where \textit{TP}, \textit{FP} and \textit{FN} denote the number of true positive, false positive and false negative elements, respectively. The $F_1$-score was defined as 
\begin{equation}
F_1-score = \frac{2\cdot precision\cdot recall}{precision+recall}.
\end{equation}
The \textit{F}-test was performed using permutational multivariate analysis of variance with sequence bootstraps \cite{Anderson2001-zz,Zrimec2018-lx}. The distribution of the \textit{F} function under the null hypothesis of no differences among group means was evaluated by performing 1e4 bootstrap repetitions, with \textit{p}-values calculated as
\begin{equation}
p = \frac{\textrm{Num. }F_{Bootstrap}\geq F}{\textrm{Total num. }F_{Bootstrap}}.
\end{equation}

\subsection{Alignment algorithm}
We developed and tested a simple ungapped DNA sequence alignment framework (Algorithm 1) that finds the most similar segments to query sequences in target sequences using a given distance function. The assessment of algorithm performance included (i) locating the transfer regions to within +/- 1 bp of their known locations in the target plasmids and (ii) correct typing of Mob groups in the target plasmids. For this, true and false positive and negative counts were obtained from the alignment tests by considering only the lowest scoring hit per alignment. A true or false positive value was assigned if the result was below a specified significance cutoff and corresponded or did not correspond, respectively, to the known value (region location or Mob group), and alternatively, a false or true negative value was assigned to results above the significance cutoff that corresponded or did not correspond, respectively, to the known value. The statistical significance of distance scores (at \textit{p}-value cutoffs from 1e-6 to 1e-1) was evaluated using bootstrap permutations (\textit{n} = 1e6 per sequence) of 10 randomly selected query sequences (Eq. 7) and a mapping function between the distance scores and \textit{p}-values was then obtained by least squares curve fitting (Matlab) to a second order polynomial function (distance score of 0 corresponded to the theoretical limit of ~1e-132) \cite{Zrimec2020-wx}. 

\begin{algorithm}
\footnotesize
\SetAlgoLined
 \textbf{input} query\_set, target\_set\;
 \For{i = 1 : size(query\_set)}{
  \For{j = 1 : size(target\_set)}{
   \For{k = 1 : length(target\_set(j))}{
    dist(i,j,k) = distance(query\_set(i),target\_set(j)(k))\;
   }
  }
 }
 \textbf{return} min(dist(:,:)).
 \caption{Sequence alignment algorithm.}
\end{algorithm}

\subsection{Software}
Matlab v2017 (www.mathworks.com), Python v3.6 (www.python.org) and R v3.5 (www.r-project.org) were used. The code is available at \url{https://www.github.com/JanZrimec/smer\_acm\_bcb\_20}.

\section{Results}
\subsection{Fusion of DNA structural properties into a compact representation encapsulates main protein-DNA binding features}
Considering that the first 4 neighboring nucleotides have the largest effect on the structural state of a given nucleotide base pair \cite{Zrimec2015-xf,Peyrard2009-as}, we designed nucleotide-position specific DNA structural representations that included the effects of 1 to 4 neighboring nucleotides, termed s-mers (Table 2: sizes 3, 5, 7 and 9 bp, Fig. 1). The s-mers were based on calculating up to 64 of the most widely used DNA structural properties for all permutations of the corresponding equally sized k-mers (Table 2, Methods M1), followed by dimensionality reduction and clustering (Fig. 1A, Methods M2). The calculated DNA structural properties included physicochemical, conformational and protein-DNA binding variables of experimental origin \cite{Zrimec2020-wx} (Table 1). 

Dimensionality reduction with Principal Component Analysis (PCA) yielded a specific number of principal components (PC) with each s-mer size \textit{s} that explained over 99\% of the data variance (Fig. 2). The amount of PCs increased with the s-mer size and with the number of initial structural variables from 14 to 18 PCs, which was an over 3.5-fold decrease in the amount of variables required to describe almost all of the original information (Table 2). On average, 3, 6, 9 and 17 components were required to explain over 60, 80, 90, and 99\% of the variance, respectively. The coefficient of variation (\(\sigma/\mu\)) of the first 6 most informative PCs across the s-mer sizes was below 0.637 and lowest with the first and sixth components with 0.160 and 0.466, respectively, showing that these PCs carried similar structural information with each \textit{s}. Analysis of PCA loadings showed that each PC mainly comprised a number of distinct structural variables, which enabled us to determine the key protein-DNA binding features defined by the 6 most informative components (in the order of decreasing importance): thermodynamic stability \cite{SantaLucia1998-hc,Protozanova2004-xc}, horizontal flexibility \cite{Olson1998-rw,Brukner1995-pt,Bolshoy1991-ux,Gorin1995-es,Karas1996-qz,Chiu2016-kb}, torsional flexibility \cite{Perez2004-sx,Kabsch1982-tv,Olson1998-rw,Karas1996-qz}, conformational stability \cite{SantaLucia1998-hc,Aida1988-iq,Perez2004-sx,Packer2000-ri,Geggier2010-mw}, major and minor groove accessibility \cite{Protozanova2004-xc,Kabsch1982-tv,Gorin1995-es,Olson1998-rw} and A-DNA to B-DNA transition potential \cite{Aida1988-iq,Perez2004-sx,Gorin1995-es,Karas1996-qz}. Moreover, the top 20 sorted structural variables contained 3 of the well known DNAshape functions \cite{Chiu2016-kb,Rohs2009-hm} and ORChID2 \cite{Bishop2011-jm}, with nucleosome positioning (phase) \cite{Satchwell1986-me} being the most important.

\begin{table*}
\footnotesize
  \caption{Summary of s-mer construction. The groups of s-mers comprised all permutations of nucleotide k-mers of 3 to 9 bps in length. Due to sequence length constraints, not all DNA structural models could be computed with all s-mers. The amount of structural models used is given under 'Structures'. The given amount of principal components (PC) describing over 99\% of the variance of the data was chosen to define each structural representation.}
  \begin{tabular}{p{1.5cm}|p{1.5cm}|p{1.5cm}|p{1.5cm}|p{1.5cm}|p{2cm}}
    \hline
    S-mer size \textit{s} & Neighbors & Permut. & Structures & PC \textgreater{} 0.99 & Number of clusters \textit{k}\\
    \hline
    3 & 1 & 64 & 57 & 14 & 2\string^2 to 2\string^6\\
    5 & 2 & 1024 & 62 & 17 & 2\string^2 to 2\string^8\\
    7 & 3 & 16384 & 64 & 18 & 2\string^2 to 2\string^8\\
    9 & 4 & 262144 & 64 & 18 & 2\string^2 to 2\string^8\\
    \hline
  \end{tabular}
\end{table*}

\begin{figure}[ht]
  \centering
  \includegraphics[width=7.5cm,keepaspectratio]{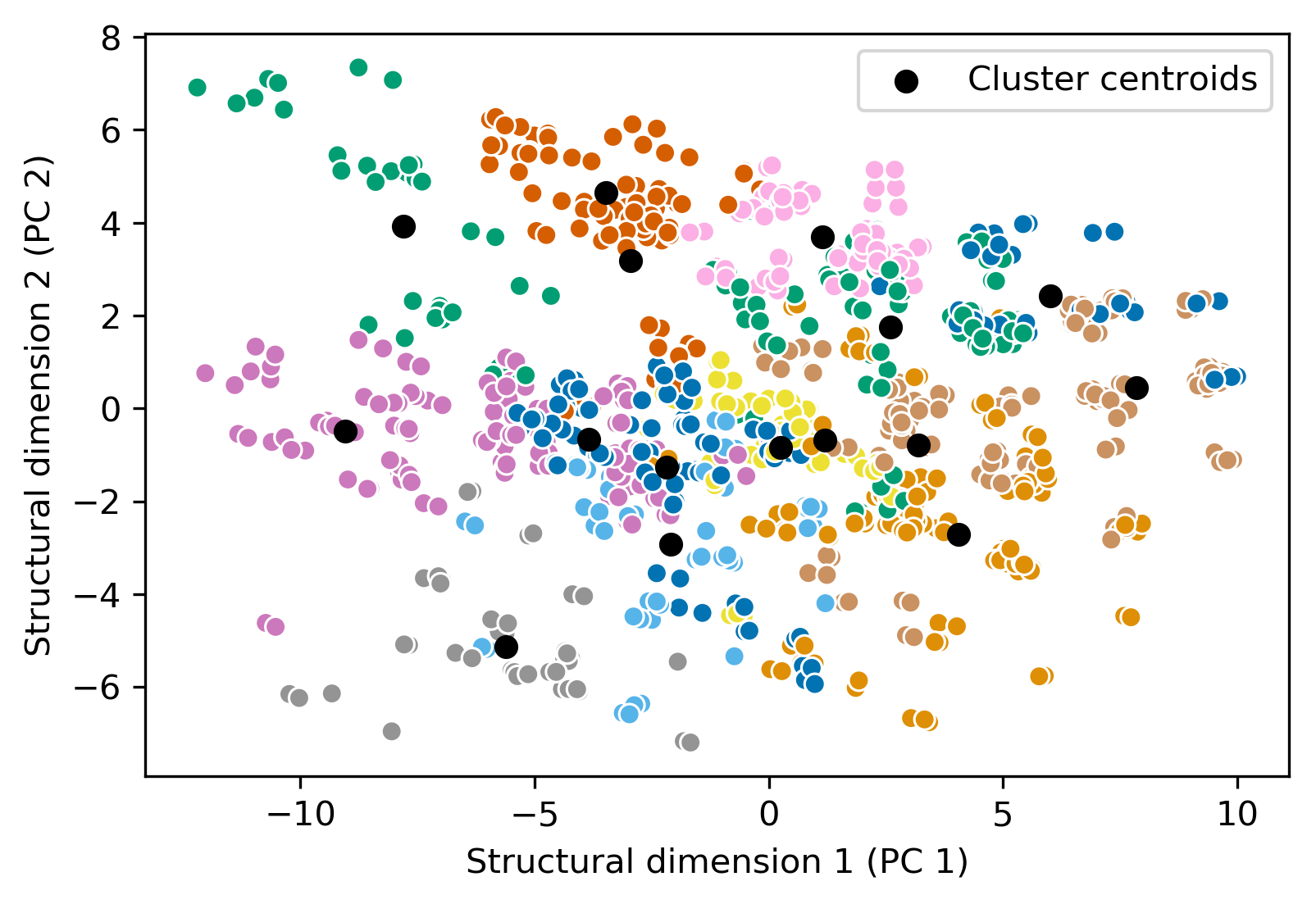}
  \caption{The first two dimensions (principal components, PC) of the structural representation with s-mer size \textit{s} = 5 and number of clusters \textit{k} = 16, where each point represents a different 5-mer nucleotide permutation \{AAAAA, AAAAC, …,TTTTT\}, colors depict the clusters and black points denote the cluster centroids.}
\end{figure}

Although the dimensionality reduced structural data was not expected to form strong clusters, we considered that a limit must exist to the resolution of the structural representations, above which there is no measurable influence on the achievable computational accuracy. Additionally, we wanted to achieve an additional level of compactness of the DNA codes as well as test the clustering of the structural data points. Clustering was performed with the number of clusters \textit{k} varied between 4 and 256 clusters (2 to 8 bits), with the exception of using up to 32 clusters with 3-mers (Table 2, Methods M2). Standard cluster evaluation methods, including Elbow and Silhouette \cite{Rousseeuw1987-mx}, showed that with a decreasing \textit{k}, the overall accuracy of the clustered data representations decreased compared to using a higher number of clusters. At the highest \textit{k} (256), the explained variance ($R^2$) was over 80\% with both s-mer sizes 5 and 7 (\textit{s} = 9 not fully tested due to memory restrictions). With a decreasing number of clusters, progressively larger clusters were obtained (Fig. 3) with a decrease in the percentage of explained variance down to ~40\% with \textit{k} = 4 clusters, and similarly a decrease in the average Silhouette ratio. The cluster sizes were approximately normally distributed (Fig. 3), with the variation of cluster sizes increasing with an increasing \textit{k},  although the coefficient of variation did not surpass 0.52.

\begin{figure}[ht]
  \centering
  \includegraphics[width=7.5cm,keepaspectratio]{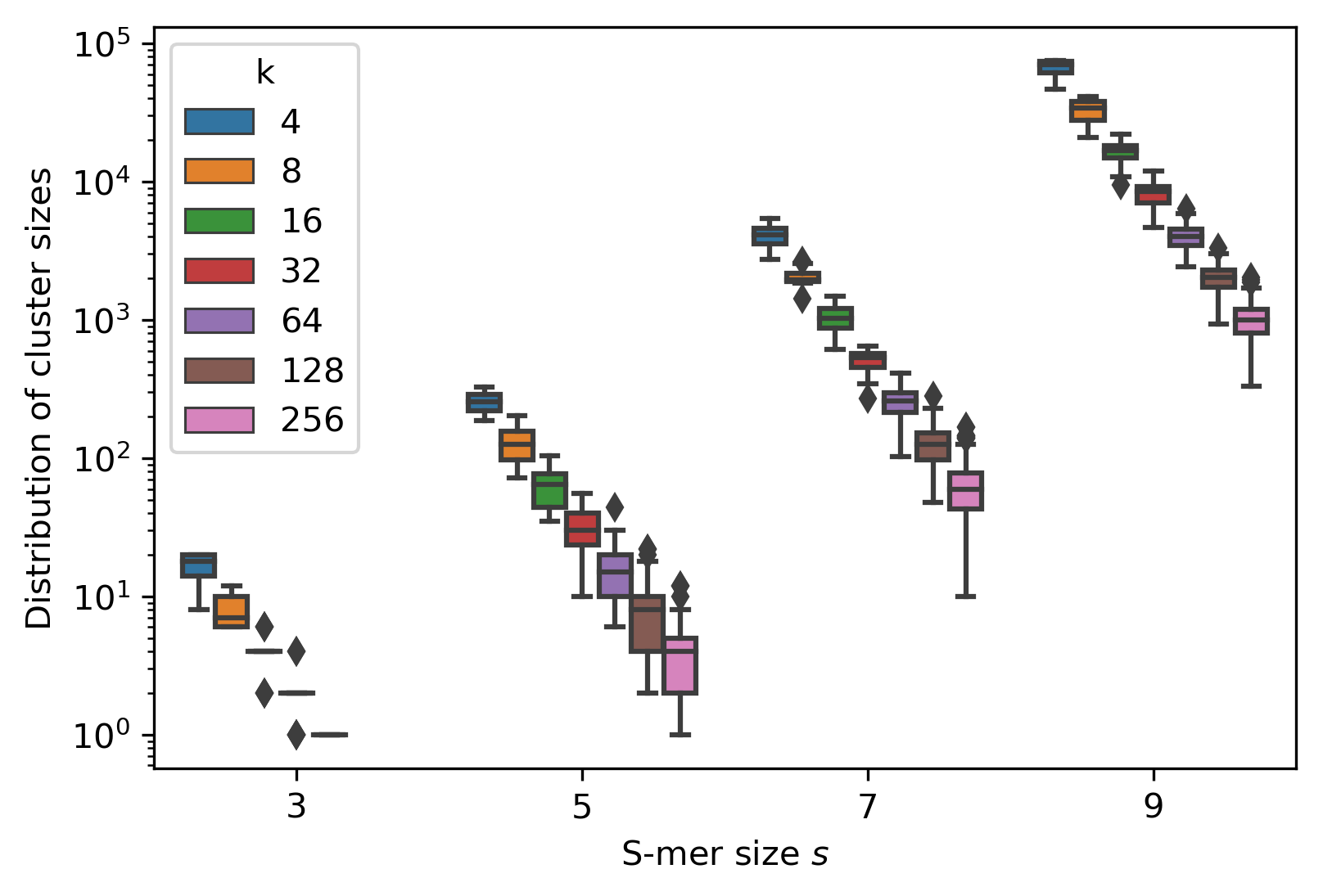}
  \caption{The distributions of cluster sizes across the structural representations with different s-mer sizes \textit{s} and number of clusters \textit{k}.}
\end{figure}

\subsection{Structural representations encode groups of functional sequence variants with conserved properties}
We next explored whether the s-mers could encode groups of conserved functional DNA motifs \cite{Levo2015-iu}, and if our structural DNA representations could more compactly but also more accurately encapsulate DNA motifs than bare k-mers. We used a dataset of 595 Jaspar \cite{Khan2018-wj} transcription factor binding site (TFBS) motifs comprising 1,296,654 unique DNA sequences from multiple model organisms (Methods M1), which contained at least 10 motifs in each group and were at least 9 bp long. To test the capacity of the structural representations to encode TFBS motifs we first measured the \textit{compression ratio} (Eq. 3), which was the ratio of the amount of unique s-mers versus the amount of unique k-mers observed with a given TFBS motif. As a note, due to computational complexity and memory limitations, only certain parameter combinations could be tested. We observed an increase in the average \textit{compression ratio} across motifs with an increasing s-mer size (Fig. 4), as it increased from 1.132 with \textit{s} = 5 (\textit{k} = 256) to 1.296 with \textit{s} = 9 (\textit{k} = 256). Similarly and as expected, the average compression ratio increased with decreasing amount of clusters (Fig. 4), from 1.132 with \textit{k} = 256 (\textit{s} = 5) to 3.152 with \textit{k} = 4 (\textit{s} = 3). The variation of the compression ratio across the different motifs was approximately constant (SD between 0.056 with \textit{s} = 3, \textit{k} = 4 and 0.092 with \textit{s} = 7, \textit{k} = 128). Moreover, we measured a significant negative correlation (Pearson's \textit{r}, \textit{p}-value \textless{} 4.7e-5) between the \textit{compression ratio} and the TFBS sequence length increasing from -0.166 to -0.667 with an increasing s-mer size \textit{s} = 3 (\textit{k} = 32) to \textit{s} = 9 (\textit{k} = 128), respectively, and up to -0.724 with a decreasing cluster size \textit{k} (s = 3, k = 4). Weak correlation (Pearson's \textit{r}, \textit{p}-value \textless{} 0.026) was also observed between the \textit{compression ratio} and the number of unique sequences in a TFBS motif, similarly as above increasing from -0.091 to -0.311 with an increasing \textit{s} = 3 (\textit{k} = 32) to \textit{s} = 9 (\textit{k} = 128), respectively, and up to -0.382 with a decreasing \textit{k} (\textit{s} = 3, \textit{k} = 4). This suggested that the capacity for compression decreases with more abundant DNA sequence space, such as with longer motifs or ones with a more diverse set of sequence variants.

\begin{figure}[ht]
  \centering
  \includegraphics[width=7.5cm,keepaspectratio]{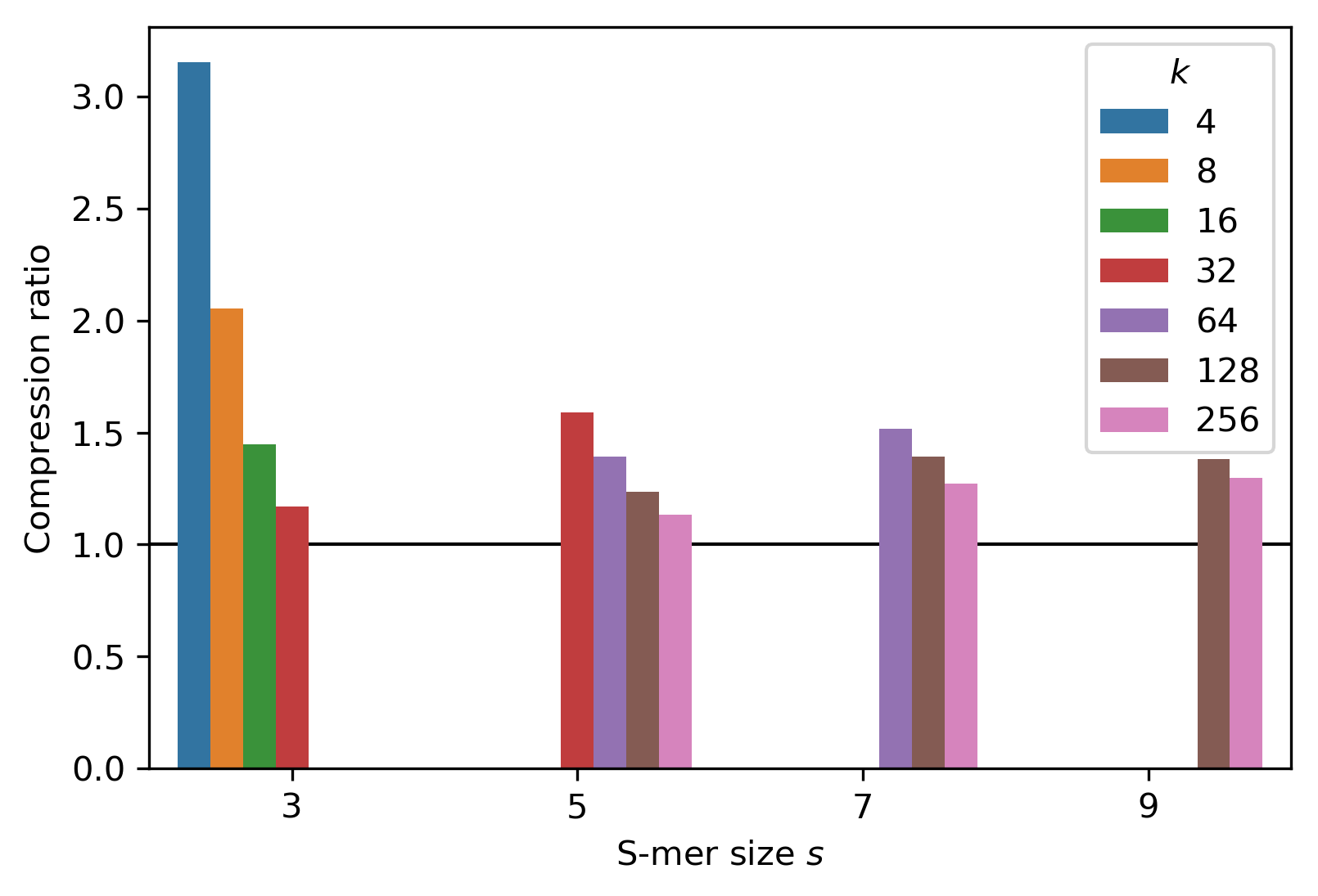}
  \caption{Compression ratios of Jaspar TFBS motifs obtained with different structural representations. Due to computation and memory limitations only certain parameter combinations of s-mer size \textit{s} and number of clusters \textit{k} were analysed. The black line denotes no compression.}
\end{figure}

Since the structural representations indeed compressed the TFBS motifs up to 3-fold, we next explored how accurate the encodings were at describing different functional motif variants. We selected the most sequence-abundant motif, the 18 bp Human MAFF motif (Jaspar: MA0495.1, class of Basic leucine zipper factors) that contained 49,462 unique sequence variants. Using a randomly selected 1\% (\textit{n} = 495) subset of these sequence variants to define their s-mers, we measured how many of the remaining 99\% of the motifs were described by (or rather, could be predicted from) these structural encodings. This meant reconstructing all the possible motif sequence variants from each structural representation instance, and gave an estimate of the encoding accuracy. The initial \textit{precision} and \textit{recall} (Methods M3) obtained without any encoding were 1 and 0.01, respectively. Unsurprisingly, an inverse relation was observed between \textit{precision} and \textit{recall} (Fig. 5). \textit{Precision} was highest (0.838) with a low s-mer size (\textit{s} = 3, \textit{k} = 32) and decreased 6.5-fold with an increasing \textit{s} (0.129, \textit{s} = 9, \textit{k} = 256), whereas \textit{recall} increased by 9\% from 0.0102 to 0.0111 at the equal parameter values, respectively. A reason for the decrease in \textit{precision} was likely that the average amount of distinct sequence variants encapsulated by the different structural representations increased with an increasing s-mer size as well as with a decreasing \textit{k} (Fig. 6). This related to an increasing coverage of the correct TFBS motifs (true positives) as well as an increasing number of variants not in the given TFBS sequence set (false positives). However, it is also possible that the TFBS sequence set is incomplete and does not contain all the possible functional sequence variants of that motif, meaning that the true positive and negative space is unknown. With an 18 bp DNA sequence, such as is the length of the MAFF motif, we estimated that up to 6.9e10 sequence variants could exist, whereas the set of sequence variants in the TFBS represented merely 7.2e-7 of this diversity and was likely undersampled. This suggested that for further analysis, datasets with a more complete coverage of the functional sequence space should be used.

\begin{figure}[ht]
  \centering
  \includegraphics[width=7.5cm,keepaspectratio]{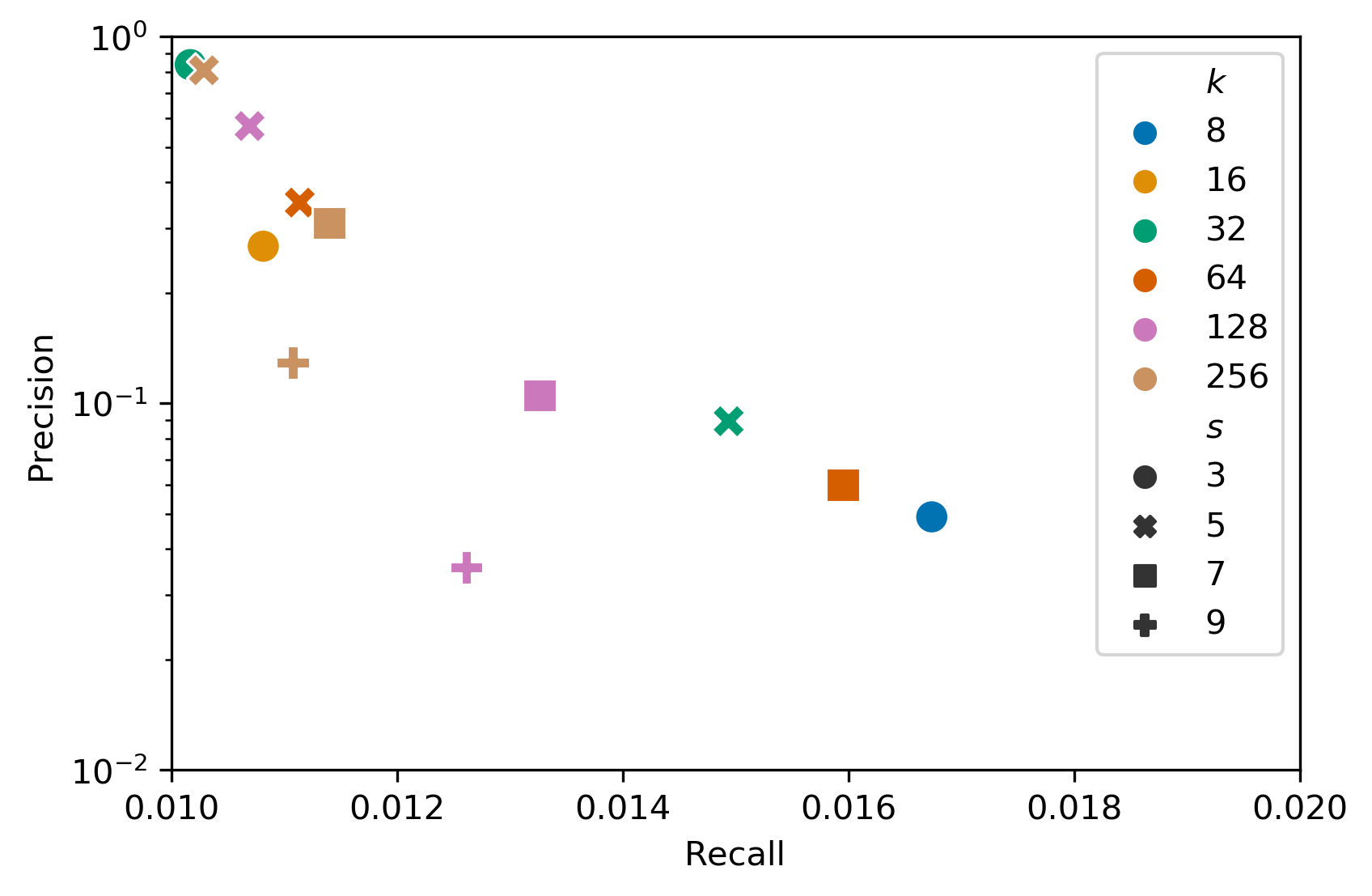}
  \caption{Precision versus recall when comparing an encoding of 1\% of sequence variants of the Human MAFF motif (Jaspar: MA0495.1) with the remaining 99\% of sequence variants, at different structural representation parameters s-mer size \textit{s} and number of clusters \textit{k}. The precision and recall obtained with nucleotide k-mers were 1 and 0.01, respectively.}
\end{figure}

\begin{figure}[ht]
  \centering
  \includegraphics[width=7.5cm,keepaspectratio]{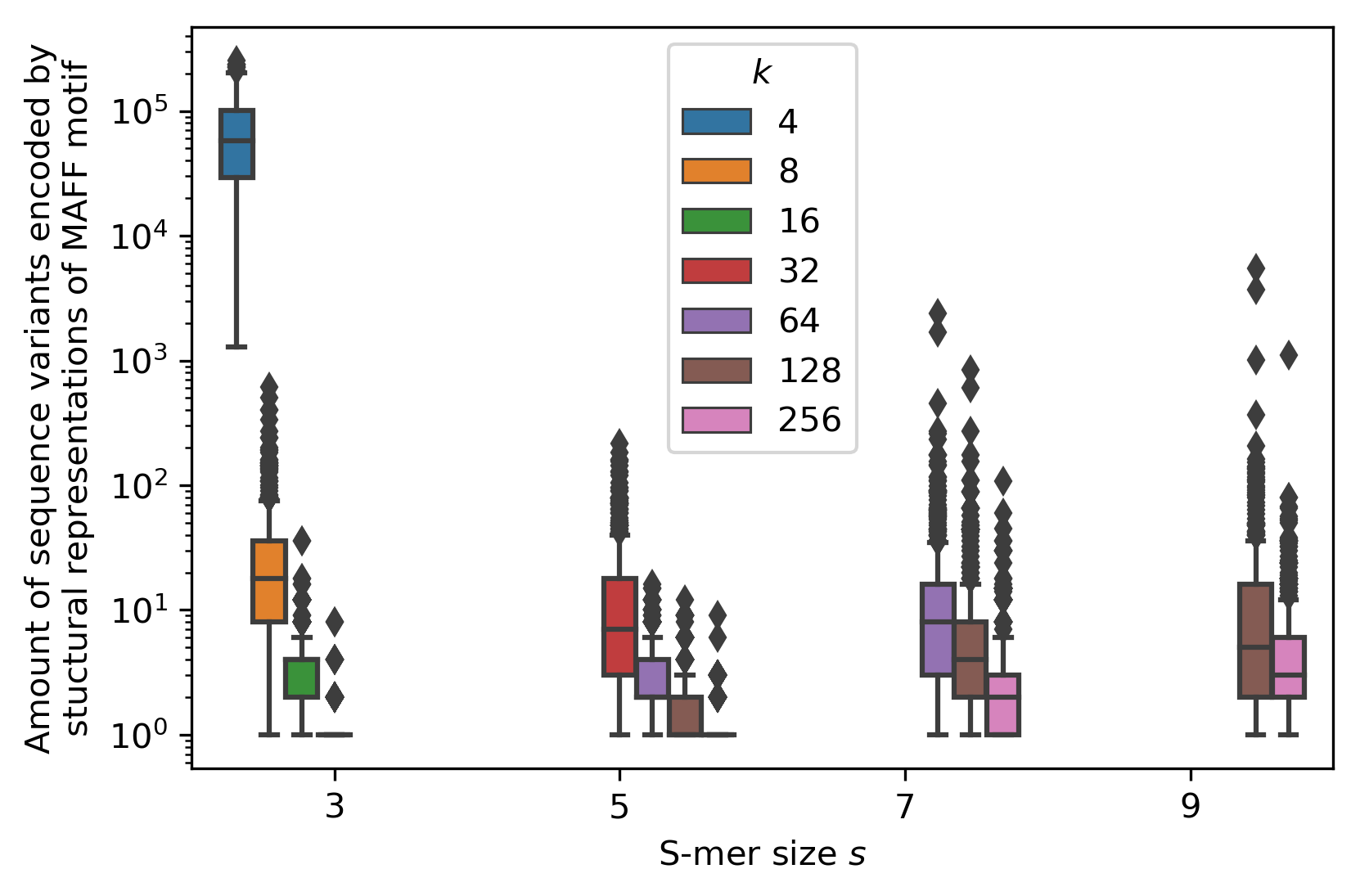}
  \caption{Amount of unique DNA sequence variants encoded by structural representations of the 18 bp long Human MAFF motif (Jaspar: MA0495.1), results with 1\% (\textit{n} = 100) of motif variants shown.}
\end{figure}

\subsection{A structure-based distance function resolves regulatory DNA more accurately than sequence-based ones}
To facilitate the comparison of two different structural representations (Fig. 1: s-mer vectors), such as is done with nucleotide sequences using for instance the p-distance (equal to the Hamming distance corrected for sequence length), we defined a structural distance, termed s-distance, as the sum of squared Euclidean distances between s-mer cluster centroids of two representations (Eq. 1). To test the s-distance and compare it to the p-distance, we used a dataset of whole DNA regulatory regions that control the initiation of DNA transfer in plasmid conjugation (origin-of-transfer regions) \cite{De_La_Cruz2010-xj} (Methods 1). These transfer regions are 220 bp long and comprise binding sites for the relaxase as well as those for accessory proteins that regulate transfer. The dataset contained 64 unique transfer sequences \cite{Zrimec2018-lx} from 4 distinct mobility groups (Mob, defined by amino acid homology of the main transfer enzyme relaxase) \cite{Garcillan-Barcia2009-yk}, with approximately equal group sizes of 16 elements. We previously determined that the Mob groups of these regions could be clearly discriminated based on 6 DNA physicochemical and conformational properties \cite{Zrimec2018-lx}. In order to additionally test the possibility of discriminating between functional and non-functional transfer sequences, here we expanded the dataset to include, besides the positive sequences (Pos) also an equal amount negative counterparts (Neg, Methods M1). Using the specific distance functions as the measure of variation across the data within a permutational MANOVA framework (bootstrap \textit{n} = 1e4, Methods M3) \cite{Anderson2005-pu}, we observed significant (\textit{p}-value \(\leq\) 1e-4) discrimination of both the positive/negative examples as well as MOB groups with the s-distance, which was not the case with the p-distance (Table 3). On average, when discriminating positive/negative examples and MOB groups with the s-distance, the amount of explained variance (R2, Eq. 2) increased 3-fold and 2.2-fold, respectively, compared to using the p-distance (Table 3). Additionally, using two datasets of Escherichia coli promoter regions, one with positive and negative examples \cite{Gusmao2014-hp} and the other grouped according to 6 most prevalent sigma factors \cite{Gama-Castro2016-so} (Methods M1), we verified the above results, as significant (\textit{p}-value \textless{} 0.05) group discrimination could be achieved with structural representations and not with the p-distance.

\begin{table*}
\footnotesize
  \caption{Comparison of distance functions based on nucleotide sequence (p-distance) and structural representations (s-distance). $F_1$-scores were obtained with the alignment algorithm (Algorithm 1).}
  \begin{tabular}{p{1.5cm}|p{1.1cm}|p{1.1cm}|p{1cm}|p{1.1cm}|p{1cm}|p{1.1cm}|p{1cm}}
  \hline
  \multirow{2}{*}{Func.} & 
  \multirow{2}{*}{Params.} & 
  \multicolumn{2}{c}{\textit{$R^2$}} & 
  \multicolumn{2}{c}{ANOVA \textit{p}-value} & 
  \multicolumn{2}{c}{$F_1$-score} \\
  \cline{3-8} & & Pos/Neg & Mob & Pos/Neg & Mob& Pos/Neg & Mob \\
  \hline
    p-distance & / & 0.017 & 0.113 & 0.472 & 0.473 & 0.892 & 0.834\\
    s-distance & \textit{s} = 3, \textit{k} = 32 & 0.046 & 0.237 & \textless{} 1e-4 & \textless{} 1e-4 & 0.955 & 0.853\\
     & \textit{s} = 5, \textit{k} = 128 & 0.049 & 0.248 & \textless{} 1e-4 & \textless{} 1e-4 & 0.942 & 0.923\\
     & \textit{s} = 7, \textit{k} = 128 & 0.054 & 0.253 & 1e-4 & \textless{} 1e-4 & 0.970 & 0.923\\
     & \textit{s} = 9, \textit{k} = 128 & 0.057 & 0.264 & 1e-4 & \textless{} 1e-4 & 0.954 & 0.879\\
  \hline
  \end{tabular}
\end{table*}

Furthermore, to determine how the amount of encoded sequence variants scales with the length of the sequence, we computed the amount of sequence variants encoded by structural representations of the 220 bp transfer regions, by randomly selecting 10 transfer regions (Fig. 7). Compared to the 18 bp TFBS motifs, structural representations of the 220 bp transfer regions encoded, on average, an over 32-fold higher number of sequence variants (Fig. 7).

\begin{figure}[ht]
  \centering
  \includegraphics[width=7.5cm,keepaspectratio]{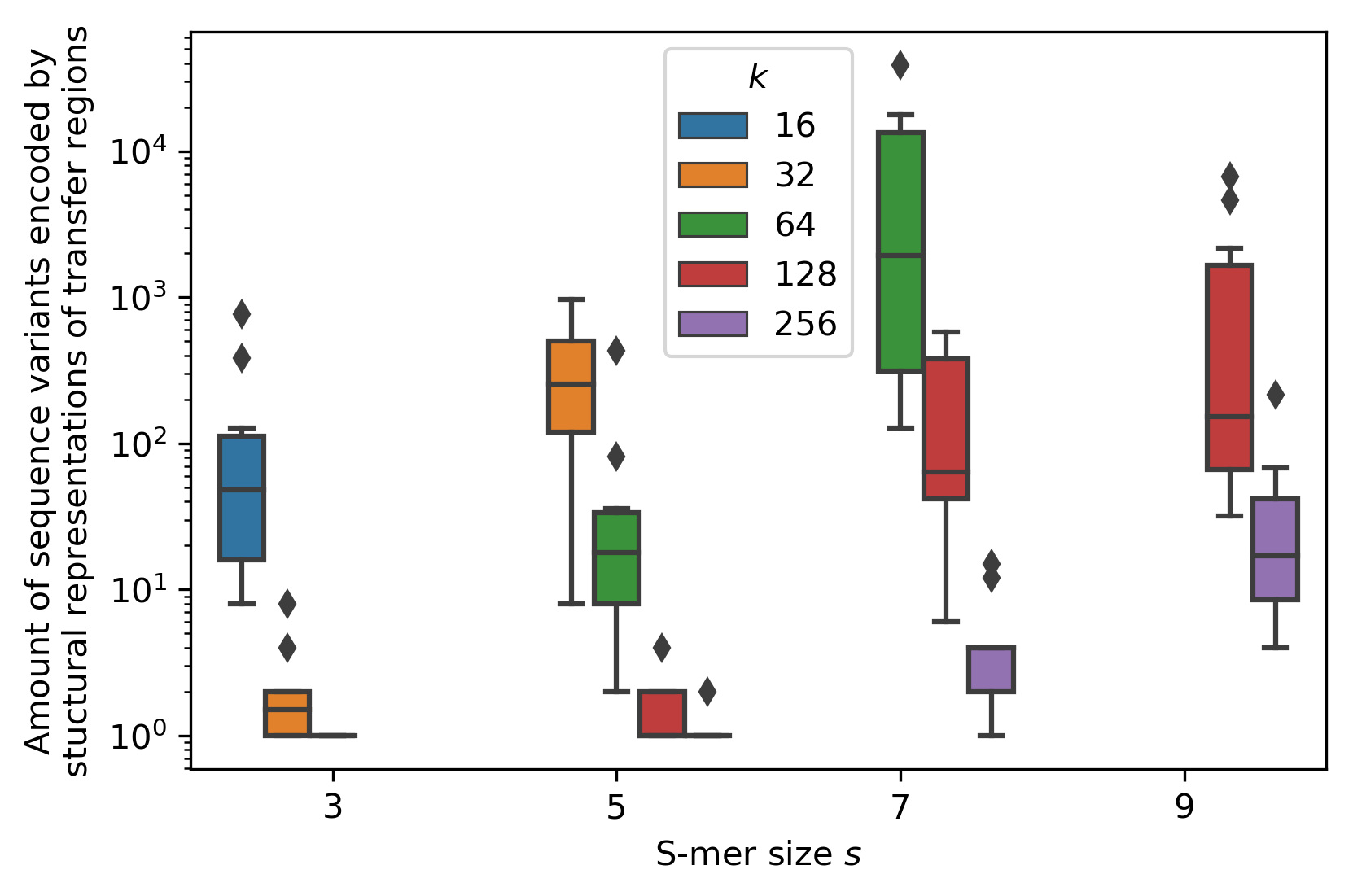}
  \caption{Amount of unique DNA sequence variants encoded by structural representations of 220 bp long plasmid transfer regions, results with 10 region variants shown.}
\end{figure}

\subsection{DNA sequence-based algorithms can be enhanced with structural representations}
We tested whether the structural representations and s-distance could be applied to existing algorithm frameworks, such as DNA sequence alignment. The DNA alignment framework that we developed (Algorithm 1) enabled the use of different metrics such as the s-distance and could align a query and a target sequence based on finding the minimal distance between them. Since equal sized groups were no longer required, we used an expanded dataset of 112 distinct transfer regions as the query dataset and a target dataset of 52 plasmids with known locations of the transfer regions as well as their Mob groups \cite{Zrimec2020-wx} (Methods M1). By counting the amount of lowest-distance alignments in the target dataset that were below a specified distance threshold, we could define similar metrics as for a binary classification problem, namely amounts of true and false positive and negative results (Methods M4) as well as the harmonic mean of precision and recall, the $F_1$-score (Eq. 6). Although both distance functions proved accurate, we observed, on average, an over 7\% improvement of the $F_1$-score with the s-distance compared to the p-distance, both when discriminating functional regions (Pos/Neg) as well as the Mob groups (\textit{p}-value \textless{} 1e-13, Table 3). The s-distance based algorithm (at \textit{s} = 7, \textit{k} = 128) thus correctly uncovered 32 (62\%) transfer regions in the target set with 30 of them (58\%) correctly Mob typed, compared to 29 (56\%) and 26 (50\%) with the sequence based p-distance, respectively. This suggested that existing DNA sequence-based algorithms could indeed be enhanced with structural representations \cite{Marcovitz2013-kg,Levo2015-iu,Slattery2014-ne,Rohs2009-hm} (Fig. 8).

\begin{figure}[ht]
  \centering
  \includegraphics[width=6cm,keepaspectratio]{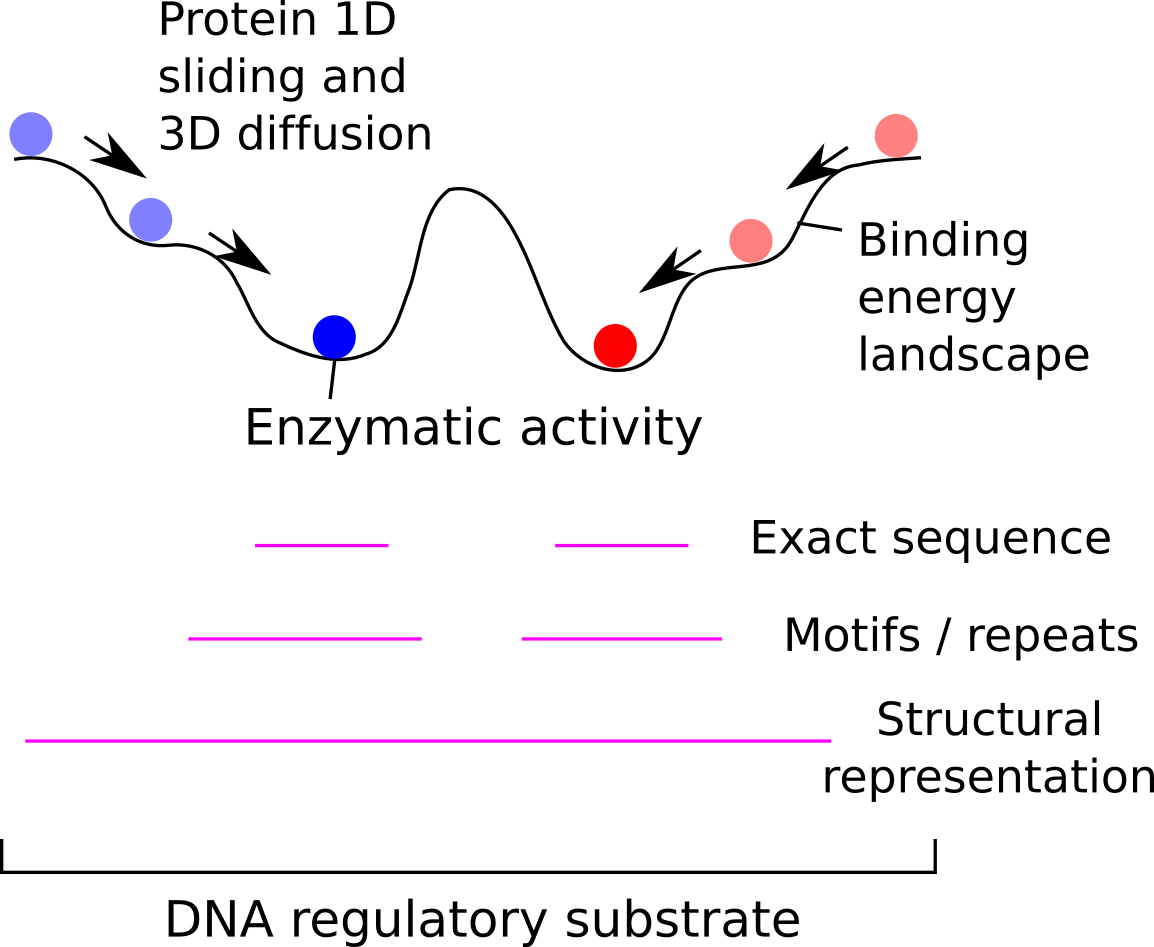}
  \caption{Possibilities for combined sequence structure algorithms based on how proteins recognize and bind their active sites in the regulatory DNA. Interactions of lower specificity with the surrounding DNA (corresponding to DNA with less conserved nucleotide sequence but defined structural properties) guide the proteins towards their specific binding sites (highly conserved or exact sequence).}
\end{figure}

\section{Discussion}
Here we used a number of functionally-relevant DNA physicochemical and conformational properties (Table 2: 57 to 64 structural variables) and fused them into compact DNA structural representations containing the most important structural information (e.g. 6 components carried over 80\% of the initial data variance). Recovery of the key distinguishing properties of the first 6 structural components showed that they indeed reflected the main properties involved in protein-DNA interactions (Table 1). The amount of structural information could be further refined with clustering down to 2 bits (Fig. 3), where the compression ratio of functional DNA sequence motifs could be as high as 3:1 (Fig. 4). Nevertheless, the most promising results were obtained with a number of clusters corresponding to 6 to 8 bits of information (64 to 256 clusters, Fig. 5). These structural representations could compress functional sequence variants by around 30\% to 50\%, where each instance of the structural representation encapsulated up to ~20 sequence variants of a TFBS motif (Fig. 6) and up to ~2000 variants of a whole DNA regulatory region (Fig. 7). Further testing is required, however, using datasets with a more complete coverage of the functional sequence space or experimentally, to properly investigate the capacity to encode groups of functional DNA sequence variants and their conserved functional properties. 

Devising the s-distance function opens up a plethora of possibilities for testing the structural codes and implementing them into DNA algorithms, as it was found to resolve regulatory DNA more accurately than sequence-based metrics \cite{Zrimec2020-wx,Zrimec2018-lx} (Table 3). One can also think of additional metrics that can prove useful, such as for instance a Jaccard distance using s-mers instead of k-mers \cite{Zrimec2020-wx}, that can distinguish coding and non-coding sequences across genomes or help bin species in metagenomic data. Furthermore, by developing and testing a distance-based alignment algorithm (Algorithm 1), we demonstrated the potential of the structural representations to enhance existing sequence-based algorithms. Here, the usefulness of sequence-like codes (similar to the nucleotide code ACGT) stood out, as the structural representations could be realized as mere sequences of cluster indices with precomputed pairwise distances, abstracting from, and altogether disposing of, the structural component space. This can simplify their implementation in existing sequence-based algorithms and also improves the accuracy of pinpointing enzymatic sites, such as nicking sites in transfer regions, down to a resolution of 1 bp \cite{Zrimec2020-wx}. Accordingly, the solution can uncover many new variants of transfer regions in natural plasmids, helping researchers investigate the potential for plasmid mobility and its global effects \cite{Zrimec2020-wx}.

The use of whole sets of structural variables is more frequently found in the domain of machine learning (ML), where ML models trained on specific sets of target variables then learn to use only specific subsets of the structural features \cite{Zrimec2015-xf,Zrimec2018-lx,Zrimec2013-ds}. For instance, when training ML models for discriminating Pos/Neg examples or MOB groups (Table 3), different variables at different locations were found to be most informative \cite{Zrimec2018-lx}. Although this leads to efficient discriminative or predictive models for specific tasks, DNA sequence analysis frequently requires generalizing across multiple tasks and using all of the features (e.g. alignment). Indeed, we have found that sequence based models of structural properties outperform ML models \cite{Zrimec2020-wx,Zrimec2018-lx}. Deep learning algorithms, however, might prove to be much more capable as they can interpret new data representations themselves \cite{Bengio2013-xg} and, in our experience, outperform any structural feature-based model using mere nucleotide sequence \cite{Zrimec2019-ql}.

Besides sequence alignment \cite{Zrimec2020-wx}, the potential uses of the structural representations include: (i) phylogenetic analysis of regulatory DNA regions \cite{Garcillan-Barcia2009-yk}, (ii) analysis of single nucleotide variations \cite{Watson2008-dt}, as the structural representations contain variants with position-specific nucleotide substitutions, (iii) motif identification \cite{Samee2019-xj}, where, for instance, the initialization stage of graph-based algorithms could be performed using structural representations \cite{Stepancic2014-yu}, and (iv) design of DNA substrates with a modified DNA sequence but conserved functionality. Other, combined approaches could potentially mimic the protein-DNA search and binding dynamics in DNA regulatory regions \cite{Marcovitz2013-kg,Levo2015-iu,Slattery2014-ne,Rohs2009-hm} and adopt a combination of both structural and sequence features (Fig. 8). Another possibility is to use different representations for the non-coding (structural) and coding (sequence) regions, e.g. for whole-gene identification. Indeed, due to the bias of the current bioinformatic methods to nucleotide sequence based approaches, it is possible that considerable performance increases might still be achievable in certain fields with such novel solutions \cite{Zrimec2020-wx, Samee2019-xj}.

\section*{Acknowledgements}
This work was supported by the Slovenian Research Agency under grant agreement no. [Z2-7257]. I kindly thank Filip Buric (Chalmers Univ. of Tech., Sweden), Tomaz Pisanski and Nino Basic (UP-FAMNIT, Slovenia), Chrats Melkonian (VU, Netherlands), Maria Pilar Garcillan-Barcia and Fernando de la Cruz (UNICAN, Spain), Joshua Ramsey (Curtin Univ., Australia) and Ziva Stepancic and Ales Lapanje (IJS, Slovenia) for their discussions and support of this research.

\bibliographystyle{unsrt}
\bibliography{2020_Zrimec-representations}

\begin{thebibliography}{10}

\bibitem{Rohs2009-hm}
Remo Rohs, Sean~M West, Alona Sosinsky, Peng Liu, Richard~S Mann, and Barry
  Honig.
\newblock The role of {DNA} shape in {protein--DNA} recognition.
\newblock {\em Nature}, 461(7268):1248--1253, October 2009.

\bibitem{Zrimec2015-xf}
Jan Zrimec and Ales Lapanje.
\newblock Fast prediction of {DNA} melting bubbles using {DNA} thermodynamic
  stability.
\newblock {\em IEEE/ACM Trans. Comput. Biol. Bioinform.}, 12(5):1137--1145,
  September 2015.

\bibitem{Zrimec2018-lx}
Jan Zrimec and Ale{\v s} Lapanje.
\newblock {DNA} structure at the plasmid origin-of-transfer indicates its
  potential transfer range.
\newblock {\em Sci. Rep.}, 8(1):1820, January 2018.

\bibitem{Chen2012-gd}
Wei Chen, Pengmian Feng, and Hao Lin.
\newblock Prediction of replication origins by calculating {DNA} structural
  properties.
\newblock {\em FEBS Lett.}, 586(6):934--938, March 2012.

\bibitem{Zrimec2020-wx}
Jan Zrimec.
\newblock Multiple plasmid origin-of-transfer substrates enable the spread of
  natural antimicrobial resistance to human pathogens.
\newblock {\em bioRxiv}, page 2020.04.20.050401, April 2020.

\bibitem{Watson2008-dt}
James~D Watson, Tania~A Baker, Stephen~P Bell, Alexander Gann, Michael~K
  Levine, and Richard Losick.
\newblock {\em Molecular Biology of the Gene. 6th. ed}.
\newblock Pearson/Benjamin Cummings, 2008.

\bibitem{De_La_Cruz2010-xj}
Fernando De~La~Cruz, Laura~S Frost, Richard~J Meyer, and Ellen~L Zechner.
\newblock Conjugative {DNA} metabolism in gram-negative bacteria.
\newblock {\em FEMS Microbiol. Rev.}, 34(1):18--40, January 2010.

\bibitem{Levo2015-iu}
Michal Levo, Einat Zalckvar, Eilon Sharon, Ana~Carolina Dantas~Machado, Yael
  Kalma, Maya Lotam-Pompan, Adina Weinberger, Zohar Yakhini, Remo Rohs, and
  Eran Segal.
\newblock Unraveling determinants of transcription factor binding outside the
  core binding site.
\newblock {\em Genome Res.}, 25(7):1018--1029, July 2015.

\bibitem{Marcovitz2013-kg}
Amir Marcovitz and Yaakov Levy.
\newblock Weak frustration regulates sliding and binding kinetics on rugged
  {Protein--DNA} landscapes.
\newblock {\em The Journal of Physical Chemistry B}, 117(42):13005--13014,
  2013.

\bibitem{Tosato2017-nq}
Valentina Tosato, Nicole West, Jan Zrimec, Dmitri~V Nikitin, Giannino Del~Sal,
  Roberto Marano, Michael Breitenbach, and Carlo~V Bruschi.
\newblock Bridge-induced translocation between {NUP145} and {TOP2} yeast genes
  models the genetic fusion between the human orthologs associated with acute
  myeloid leukemia.
\newblock {\em Front. Oncol.}, 7:231, 2017.

\bibitem{Slattery2014-ne}
Matthew Slattery, Tianyin Zhou, Lin Yang, Ana~Carolina Dantas~Machado, Raluca
  Gord{\^a}n, and Remo Rohs.
\newblock Absence of a simple code: how transcription factors read the genome.
\newblock {\em Trends Biochem. Sci.}, 39(9):381--399, September 2014.

\bibitem{SantaLucia1998-hc}
J~SantaLucia, Jr.
\newblock A unified view of polymer, dumbbell, and oligonucleotide {DNA}
  nearest-neighbor thermodynamics.
\newblock {\em Proc. Natl. Acad. Sci. U. S. A.}, 95(4):1460--1465, February
  1998.

\bibitem{Bishop2011-jm}
Eric~P Bishop, Remo Rohs, Stephen C~J Parker, Sean~M West, Peng Liu, Richard~S
  Mann, Barry Honig, and Thomas~D Tullius.
\newblock A map of minor groove shape and electrostatic potential from hydroxyl
  radical cleavage patterns of {DNA}.
\newblock {\em ACS Chem. Biol.}, 6(12):1314--1320, December 2011.

\bibitem{Chiu2016-kb}
Tsu-Pei Chiu, Federico Comoglio, Tianyin Zhou, Lin Yang, Renato Paro, and Remo
  Rohs.
\newblock {DNAshapeR}: an {R/Bioconductor} package for {DNA} shape prediction
  and feature encoding.
\newblock {\em Bioinformatics}, 32(8):1211--1213, April 2016.

\bibitem{Brukner1995-pt}
Ivan Brukner, Roberto Sanchez, Dietrich Suck, and Sandor Pongor.
\newblock Sequence-dependent bending propensity of {DNA} as revealed by {DNase}
  i: parameters for trinucleotides.
\newblock {\em EMBO J.}, 14(8):1812--1818, 1995.

\bibitem{Geggier2010-mw}
Stephanie Geggier and Alexander Vologodskii.
\newblock Sequence dependence of {DNA} bending rigidity.
\newblock {\em Proc. Natl. Acad. Sci. U. S. A.}, 107(35):15421--15426, August
  2010.

\bibitem{Karas1996-qz}
H~Karas, R~Kn{\"u}ppel, W~Schulz, H~Sklenar, and E~Wingender.
\newblock Combining structural analysis of {DNA} with search routines for the
  detection of transcription regulatory elements.
\newblock {\em Comput. Appl. Biosci.}, 12(5):441--446, October 1996.

\bibitem{Olson1998-rw}
Wilma~K Olson, Andrey~A Gorin, Xiang-Jun Lu, Lynette~M Hock, and Victor~B
  Zhurkin.
\newblock {DNA} sequence-dependent deformability deduced from {protein--DNA}
  crystal complexes.
\newblock {\em Proc. Natl. Acad. Sci. U. S. A.}, 95(19):11163--11168, September
  1998.

\bibitem{Perez2004-sx}
Alberto Perez, Agnes Noy, Filip Lankas, F~Javier Luque, and Modesto Orozco.
\newblock The relative flexibility of {B-DNA} and {A-RNA} duplexes: database
  analysis.
\newblock {\em Nucleic Acids Res.}, 32(20):6144--6151, 2004.

\bibitem{Aida1988-iq}
Misako Aida.
\newblock An ab initio molecular orbital study on the sequence-dependency of
  {DNA} conformation: An evaluation of intra- and inter-strand stacking
  interaction energy.
\newblock {\em Journal of Theoretical Biology}, 130(3):327--335, 1988.

\bibitem{Hartmann1989-ji}
B~Hartmann, B~Malfoy, and R~Lavery.
\newblock Theoretical prediction of base sequence effects in {DNA}.
  experimental reactivity of {Z-DNA} and {B-Z} transition enthalpies.
\newblock {\em J. Mol. Biol.}, 207(2):433--444, May 1989.

\bibitem{Kulkarni2013-xm}
Mandar Kulkarni and Arnab Mukherjee.
\newblock Sequence dependent free energy profiles of localized {B-} to a-form
  transition of {DNA} in water.
\newblock {\em J. Chem. Phys.}, 139(15):155102, October 2013.

\bibitem{Ho1986-hg}
Pui~S Ho, Michael~J Ellison, Gary~J Quigley, and Alexander Rich.
\newblock A computer aided thermodynamic approach for predicting the formation
  of {Z-DNA} in naturally occurring sequences.
\newblock {\em EMBO J.}, 5(10):2737--2744, 1986.

\bibitem{Zrimec2013-ds}
Jan Zrimec, Rok Kopin{\v c}, Toma{\v z} Rijavec, Tatjana Zrimec, and Ale{\v s}
  Lapanje.
\newblock Band smearing of {PCR} amplified bacterial {16S} {rRNA} genes:
  dependence on initial {PCR} target diversity.
\newblock {\em J. Microbiol. Methods}, 95(2):186--194, November 2013.

\bibitem{Samee2019-xj}
Md~Abul~Hassan Samee, Benoit~G Bruneau, and Katherine~S Pollard.
\newblock A de novo shape motif discovery algorithm reveals preferences of
  transcription factors for {DNA} shape beyond sequence motifs.
\newblock {\em Cell Systems}, 8(1):27--42.e6, 2019.

\bibitem{Bansal2014-ko}
Manju Bansal, Aditya Kumar, and Venkata~Rajesh Yella.
\newblock Role of {DNA} sequence based structural features of promoters in
  transcription initiation and gene expression.
\newblock {\em Curr. Opin. Struct. Biol.}, 25:77--85, April 2014.

\bibitem{Lucas2010-gi}
Mar{\'\i}a Lucas, Blanca Gonz{\'a}lez-P{\'e}rez, Matilde Cabezas, Gabriel
  Moncalian, Germ{\'a}n Rivas, and Fernando de~la Cruz.
\newblock Relaxase {DNA} binding and cleavage are two distinguishable steps in
  conjugative {DNA} processing that involve different sequence elements of the
  nic site.
\newblock {\em J. Biol. Chem.}, 285(12):8918--8926, March 2010.

\bibitem{Sut2009-kg}
Marta~V Sut, Sanja Mihajlovic, Silvia Lang, Christian~J Gruber, and Ellen~L
  Zechner.
\newblock Protein and {DNA} effectors control the {TraI} conjugative helicase
  of plasmid {R1}.
\newblock {\em J. Bacteriol.}, 191(22):6888--6899, November 2009.

\bibitem{Moncalian1999-qj}
Gabriel Moncalian, Mikel Valle, Jose~Maria Valpuesta, and Fernando de~la Cruz.
\newblock {IHF} protein inhibits cleavage but not assembly of plasmid {R388}
  relaxosomes.
\newblock {\em Molecular Microbiology}, 31(6):1643--1652, 1999.

\bibitem{Williams2007-be}
Sarah~L Williams and Joel~F Schildbach.
\newblock {TraY} and integration host factor orit binding sites and {F}
  conjugal transfer: sequence variations, but not altered spacing, are
  tolerated.
\newblock {\em J. Bacteriol.}, 189(10):3813--3823, May 2007.

\bibitem{Khan2018-wj}
Aziz Khan, Oriol Fornes, Arnaud Stigliani, Marius Gheorghe, Jaime~A
  Castro-Mondragon, Robin van~der Lee, Adrien Bessy, Jeanne Ch{\`e}neby,
  Shubhada~R Kulkarni, Ge~Tan, Damir Baranasic, David~J Arenillas, Albin
  Sandelin, Klaas Vandepoele, Boris Lenhard, Beno{\^\i}t Ballester, Wyeth~W
  Wasserman, Fran{\c c}ois Parcy, and Anthony Mathelier.
\newblock {JASPAR} 2018: update of the open-access database of transcription
  factor binding profiles and its web framework.
\newblock {\em Nucleic Acids Res.}, 46(D1):D1284, January 2018.

\bibitem{Gusmao2014-hp}
E~G Gusm{\"a}o and M~C~P de~Souto.
\newblock Issues on sampling negative examples for predicting prokaryotic
  promoters.
\newblock In {\em 2014 International Joint Conference on Neural Networks
  ({IJCNN})}, pages 494--501. IEEE, July 2014.

\bibitem{Gama-Castro2016-so}
Socorro Gama-Castro, Heladia Salgado, Alberto Santos-Zavaleta, Daniela
  Ledezma-Tejeida, Luis Mu{\~n}iz-Rascado, Jair~Santiago Garc{\'\i}a-Sotelo,
  Kevin Alquicira-Hern{\'a}ndez, Irma Mart{\'\i}nez-Flores, Lucia Pannier,
  Jaime~Abraham Castro-Mondrag{\'o}n, Alejandra Medina-Rivera, Hilda
  Solano-Lira, C{\'e}sar Bonavides-Mart{\'\i}nez, Ernesto P{\'e}rez-Rueda,
  Shirley Alquicira-Hern{\'a}ndez, Liliana Porr{\'o}n-Sotelo, Alejandra
  L{\'o}pez-Fuentes, Anastasia Hern{\'a}ndez-Koutoucheva, V{\'\i}ctor Del
  Moral-Ch{\'a}vez, Fabio Rinaldi, and Julio Collado-Vides.
\newblock {RegulonDB} version 9.0: high-level integration of gene regulation,
  coexpression, motif clustering and beyond.
\newblock {\em Nucleic Acids Res.}, 44(D1):D133--43, January 2016.

\bibitem{Rousseeuw1987-mx}
Peter~J Rousseeuw.
\newblock Silhouettes: a graphical aid to the interpretation and validation of
  cluster analysis.
\newblock {\em J. Comput. Appl. Math.}, 20:53--65, 1987.

\bibitem{Tibshirani2001-wq}
Robert Tibshirani, Guenther Walther, and Trevor Hastie.
\newblock Estimating the number of clusters in a data set via the gap
  statistic.
\newblock {\em J. R. Stat. Soc. Series B Stat. Methodol.}, 63(2):411--423, May
  2001.

\bibitem{Schneider1986-iv}
T~D Schneider, G~D Stormo, L~Gold, and A~Ehrenfeucht.
\newblock Information content of binding sites on nucleotide sequences.
\newblock {\em J. Mol. Biol.}, 188(3):415--431, April 1986.

\bibitem{Anderson2001-zz}
Marti~J Anderson.
\newblock A new method for non-parametric multivariate analysis of variance.
\newblock {\em Austral Ecol.}, 26(1):32--46, 2001.

\bibitem{Peyrard2009-as}
M~Peyrard, S~Cuesta-L{\'o}pez, and D~Angelov.
\newblock Experimental and theoretical studies of sequence effects on the
  fluctuation and melting of short {DNA} molecules.
\newblock {\em J. Phys. Condens. Matter}, 21(3):034103, January 2009.

\bibitem{Protozanova2004-xc}
Ekaterina Protozanova, Peter Yakovchuk, and Maxim~D Frank-Kamenetskii.
\newblock {Stacked--Unstacked} equilibrium at the nick site of {DNA}.
\newblock {\em Journal of Molecular Biology}, 342(3):775--785, 2004.

\bibitem{Bolshoy1991-ux}
A~Bolshoy, P~McNamara, and {others}.
\newblock Curved {DNA} without {AA}: experimental estimation of all 16 {DNA}
  wedge angles.
\newblock {\em Proceedings of the}, 6(88):2312--6, 1991.

\bibitem{Gorin1995-es}
A~A Gorin, V~B Zhurkin, and W~K Olson.
\newblock {B-DNA} twisting correlates with base-pair morphology.
\newblock {\em J. Mol. Biol.}, 247(1):34--48, March 1995.

\bibitem{Kabsch1982-tv}
W~Kabsch, C~Sander, and E~N Trifonov.
\newblock The ten helical twist angles of {B-DNA}.
\newblock {\em Nucleic Acids Research}, 10(3):1097--1104, 1982.

\bibitem{Packer2000-ri}
M~J Packer, M~P Dauncey, and C~A Hunter.
\newblock Sequence-dependent {DNA} structure: dinucleotide conformational maps.
\newblock {\em J. Mol. Biol.}, 295(1):71--83, January 2000.

\bibitem{Satchwell1986-me}
S~C Satchwell, H~R Drew, and A~A Travers.
\newblock Sequence periodicities in chicken nucleosome core {DNA}.
\newblock {\em J. Mol. Biol.}, 191(4):659--675, October 1986.

\bibitem{Garcillan-Barcia2009-yk}
Mar{\'\i}a~Pilar Garcill{\'a}n-Barcia, Mar{\'\i}a~Victoria Francia, and
  Fernando de~la Cruz.
\newblock The diversity of conjugative relaxases and its application in plasmid
  classification.
\newblock {\em FEMS Microbiol. Rev.}, 33(3):657--687, May 2009.

\bibitem{Anderson2005-pu}
Marti~J Anderson.
\newblock Permutational multivariate analysis of variance.
\newblock {\em Department of Statistics, University of Auckland, Auckland},
  26:32--46, 2005.

\bibitem{Bengio2013-xg}
Yoshua Bengio, Aaron Courville, and Pascal Vincent.
\newblock Representation learning: a review and new perspectives.
\newblock {\em IEEE Trans. Pattern Anal. Mach. Intell.}, 35(8):1798--1828,
  August 2013.

\bibitem{Zrimec2019-ql}
J~Zrimec, F~Buric, A~S Muhammad, R~Chen, V~Verendel, and {others}.
\newblock Gene expression is encoded in all parts of a co-evolving interacting
  gene regulatory structure.
\newblock {\em bioRxiv}, page 10.1101/792531, 2019.

\bibitem{Stepancic2014-yu}
{\v Z}iva Stepan{\v c}i{\v c}.
\newblock Enhancing gibbs sampling method for motif finding in {DNA} with
  initial graph representation of sequences.
\newblock {\em Journal of Computational Biology}, 21(10):741--752, 2014.

\end{thebibliography}

\end{document}